\documentclass[10pt,journal]{IEEEtran}
\usepackage{booktabs}
\usepackage{amsmath,amssymb,amsfonts}
\usepackage[algo2e]{algorithm2e}
\usepackage{algorithmic}
\usepackage{algorithm}
\usepackage{xcolor}
\usepackage{array}
\usepackage{textcomp}
\usepackage{stfloats}
\usepackage{url}
\usepackage{verbatim}
\usepackage{graphicx}
\usepackage{cite}
\newtheorem{pro}{Proposition}
\begin{document}

\title{An OTFS-based Random Access Scheme for GNSS Independent Operation in NTN}


\author{Marius Caus,~\IEEEmembership{Senior Member,~IEEE,} and Musbah Shaat,~\IEEEmembership{Senior Member,~IEEE,}
\thanks{M. Caus and M. Shaat are with the Centre Tecnològic de Telecomunicacions de Catalunya-CERCA (CTTC-CERCA), Castelldefels, Barcelona, Spain (email: mcaus@cttc.es; mshaat@cttc.es) }
}



\maketitle

\begin{abstract}
This paper investigates the random access procedure for non-terrestrial networks operating without global navigation satellite system (GNSS) support. In such scenarios, positioning uncertainties can reach several kilometers, which directly impacts the open-loop compensation mechanisms employed by the user equipment. To ensure that the resulting time and carrier frequency offsets can be handled by the network, the robustness of the standardized random access signal design and detection scheme must be enhanced. To extend radio access capabilities, identical Zadoff Chu (ZC) sequences are concatenated and then modulated into the orthogonal time frequency space (OTFS) modulation. Thanks to the specific characteristics of the OTFS-based random access signal, the received sequences are coherently combined, thereby maximizing the desired signal strength. Additionally, the proposed preamble minimizes the overhead associated with the cyclic prefix (CP) transmission. Numerical evaluations in a regenerative low-Earth-orbit (LEO) satellite scenario show that, despite significant positioning errors, the proposed OTFS random access design attains comparable peak-to-average power ratio (PAPR) and missed detection probability (MDP) to OFDM-based solutions, while improving spectral confinement and reducing overhead. These results demonstrate that the proposed OTFS-based random access design offers a robust and spectrally efficient alternative to OFDM for GNSS-independent NTN access.
\end{abstract}


\section{Introduction}
\label{secI}
\IEEEPARstart{N}{on-terrestrial} network (NTN) deployments are crucial to deliver broadband services in regions beyond
the reach of terrestrial infrastructures. Key use cases that can benefit significantly from the NTN connectivity include maritime environments, public protection and disaster relief (PPDR) scenarios, as well as rural and remote regions lacking cellular coverage. 

From a standardization perspective, 3GPP Release 17 defined for the first time the technical specifications of the NTN component. This release supports transparent architectures to operate in frequency range 1 (FR1) bands. The fundamental principle of NTN in 3GPP Release 17 is based on reusing legacy procedures to the highest possible extent. As part of the requirements, the user equipment (UE) shall be capable of autonomously compensating the propagation delay and the Doppler effects induced by satellite orbital motion. To achieve this, the UE must be aware of both the satellite trajectory and its own location, as specified in \cite{nr300}. To acquire the position and the velocity of the satellite, the standard has defined the system information block 19 (SIB19), which is broadcast in the downlink frames. Relying on the global navigation satellite system (GNSS), the UE is able to obtain positioning information. Remarkably, NTN has evolved during the last years and new features have been included in 3GPP Release 18. The major enhancements include the specification of deployments in the Ka band. Ongoing work in 3GPP Release 19 targets the specification of regenerative payloads, coverage enhancements, broadcasting services, deployments in the Ku band, and reduced-capability UEs. 

Looking ahead, 3GPP is preparing to standardize the 6G technology. As part of this effort, Release 20 will include a study phase focused on identifying potential 6G use cases and key features. The outcomes of these studies will feed the specification work, which is expected to begin in Release 21. While the scope of future radio technologies remains exploratory, the 6G air interface may not be backward-compatible to unleash the full potential of a new mobile generation. Innovations in the satellite radio interface are likely to include the direct access via the use of smartphones, GNSS-independent operation, the adoption of Doppler-resilient waveforms, and the use of multi-connectivity technologies, to mention a few. 

In this context, this work envisions a 6G NTN-capable UE that is able to attach to the network without GNSS support, which challenges the assumption made so far in 5G. This capability represents a significant advance to mitigate the impact of GNSS signal disruptions, which has motivated 3GPP to define a 6G use case on resilient satellite-based positioning, as detailed in \cite{nr870}. Hence, GNSS-independent positioning solutions are expected to be an integral part of 6G NTN communications. In alignment with this vision, studies such as \cite{Wan21,Zhu24} show that UEs can achieve self-positioning by leveraging time and frequency offset measurements on reference signals broadcast by NTN nodes. The accuracy of these self-positioning methods depend on the measurement time and the number of visible satellites. When only a single satellite is visible, the resulting location estimate is coarse, often with uncertainties on the order of kilometers. Such uncertainty impacts the open-loop compensation mechanism implemented by the UE. If the compensation is inaccurate, the preamble transmitted in the first step of the random access procedure (RAP) will be time and frequency misaligned at reception. Depending on the magnitude of the user positioning error, the resulting time offset (TO) and carrier frequency offset (CFO) may exceed the values tolerated by the preamble detector. This issue is especially severe in low-Earth orbit (LEO) constellations and high-frequency bands, where the Doppler effects are more prominent. To accommodate the offsets that are inherent to LEO satellite communication systems, the robustness of the random access scheme shall be enhanced. 

\subsection{Related works}
The standardized random access signal that is specified in \cite{nr211} consists of a cyclic prefix (CP) and a preamble, which is modulated into the discrete Fourier transform spread orthogonal frequency division multiplexing (DFT-s-OFDM) waveform. The preamble is generated by concatenating identical Zadoff Chu (ZC) sequences to facilitate detection in the low signal-to-noise ratio (SNR) regime. The CP is used to achieve a circular structure and prevent the interference from previous slots. Remarkably, the CP is dimensioned according to the maximum cell radius and the delay spread. In NTN, the CP is used to handle the residual TO that is induced by imperfect compensation. Reliable preamble detection requires the TO and the CFO to be less than the CP duration and half the subcarrier spacing (SCS), respectively. Otherwise, pseudo-correlation peaks arise, leading to timing ambiguities and ultimately to erroneous detection. The preamble formats defined in the standard offer different tolerance levels for time and frequency misalignments. A large SCS is beneficial to mitigate CFO effects, but it reduces CP duration, making the system more sensitive to TO. As the authors have shown in \cite{Cau24}, the available formats are not able to deal with the time and the frequency misalignment that result from an erroneous compensation in LEO satellite systems.   

The main challenge in extending the radio access capabilities stems from the trade-off between TO and CFO detection ranges. Previous studies have explored two different approaches to achieve a balanced solution. The efforts focus on enhancing detection algorithms and preamble formats to mitigate the adverse effects of: i) large CFO in configurations with narrow SCS, or ii) large TO in configurations with wide SCS. These schemes rely on the fact that small and high SCSs are able to handle large TO and large CFO values, respectively.

Several solutions have been proposed to enhance robustness against CFO with a reduced SCS, such as those in \cite{Cui15,Hua19,Li25,Zha19,Zha21,Kha21,Zhe20,Zhe21,Che22}. The superposition of ZC sequences with different root indices is an effective approach to combat the adverse effects of CFO, as discussed in \cite{Cui15,Hua19,Li25}. Another promising method to eliminate the impact of CFO is the concatenation of ZC sequences with multiple roots, as shown in \cite{Zha19,Zha21,Kha21}. The works in \cite{Zhe20,Zhe21} introduce a detection algorithm that is designed to mitigate CFO impairments, by using the differential correlation metric. However, it is important to remark that the detector exhibits a trade-off between complexity and timing estimation performance. In \cite{Che22}, the detector counteracts the CFO by performing multiple hypotheses of discrete CFO candidates with a resolution of 1.25 kHz. The main drawback is the limited capacity for scalability, as the number of hypotheses grows with the magnitude of the CFO, which has an impact on the complexity of the detector. The schemes presented in \cite{Cui15,Hua19,Li25,Zha19,Zha21,Kha21,Zhe20,Zhe21,Che22} resolve timing and frequency ambiguities by using multiple root ZC sequences for preamble generation. This, however, reduces the overall efficiency of the preamble generation method, as well as the random access capacity.

The common approach to extend the TO detection range is based on cascading ZC sequences and implementing two-step detection algorithms. The aim of employing two detection steps is to sequentially estimate fractional and integer delays. Typically, fractional and integer delay components are computed with respect to the length of a single ZC sequence. The rationale behind this factorization is to reduce the search space. 
Upon performing the detection, the total delay can be estimated, taking into account the compound effects of fractional and integer delays. It is important to note that the two-step detection approach inherently suffers from uneven performance between fractional and integer delay estimation,
as well as from error propagation between the two stages.

In the literature, several preamble designs implement the two-step detection scheme to achieve a large TO detection range, e.g., \cite{Zhe18,Cau24,Che24,Cau24b,Zhu24}. The solutions proposed in \cite{Zhe18, Cau24} are based on concatenating a single root ZC sequence with various cyclic shifts. In contrast, the method in \cite{Che24} interleaves two ZC sequences, which could be generated from the same ZC root index, following a predefined pattern. Remarkably, the solutions addressed in \cite{Zhe18,Cau24,Che24} depart from the standardized preamble generation method described in \cite{nr211}. To minimize the impact on the standard, the authors in \cite{Cau24b} have shown the effectiveness of repeating a single root ZC sequence with a single shift in estimating large TOs. An improved preamble design is presented in \cite{Zhu24}, which outperforms the schemes in \cite{Zhe18,Cau24,Che24,Cau24b} by eliminating the CP. Nevertheless, this preamble is constructed by using two ZC root sequences. Although this design reduces overhead, it limits the number of available ZC sequences and, consequently, reduces the random access capacity. As previously discussed, this limitation is also observed in \cite{Cui15,Hua19,Li25,Zha19,Zha21,Kha21,Zhe20,Zhe21,Che22}.

Recently, some works have emphasized the importance of the waveform design on the random access scheme, e.g., \cite{Cau20,Cau23,Sin20}. In \cite{Cau20,Cau23}, the superior performance of the filterbank multicarrier modulation (FBMC) over DFT-s-OFDM is shown. In this regard, FBMC exhibits much lower out-of-band (OOB) emissions and achieves lower missed detection probability (MDP) in presence of CFO. Nonetheless, the detector implemented in \cite{Cau20} is very demanding in terms of
computational effort. The complexity burden comes from the sliding window approach, which is adopted to search for all possible TO values. The complexity is reduced in \cite{Cau23}, but the resulting scheme provides similar TO and CFO detection ranges as the standard preamble formats specified in \cite{nr211}. In contrast, the solution devised in \cite{Sin20} resorts to the orthogonal time and frequency space (OTFS) modulation to allocate high-power pilots in the delay-Doppler (DD) domain. The detector estimates the delay from the pilot shift in the delay domain. To enable multi-user access, the pilots shall be transmitted in different Doppler bins. By separating the pilots according to the maximum frequency shift, the OTFS-based random access offers high robustness to the CFO. However, in the presence of large TO and CFO values, the number of available access patterns becomes limited. This constraint can significantly increase the probability of collision events, resulting in performance degradation.

Despite the remarkable features of FBMC and OTFS, the schemes devised in \cite{Cau20,Cau23,Sin20} exhibit limitations when applied to GNSS-free random access systems, especially to accommodate large TO and CFO. 

To the best of our knowledge, existing preamble designs address large TO and CFO by either appending long CP
blocks or employing multiple ZC root sequences. This observation highlights that a key challenge for future designs is the development of preamble structures that can effectively mitigate these impairments, which are inherent to the GNSS-independent operation, by using only a single ZC root sequence and minimal CP overhead. 

\subsection{Main contributions}
The enhancement proposed in this paper is based on modulating ZC sequences into OTFS. The use of OTFS is endorsed by its effectiveness in achieving reliable communications in doubly-selective channels, as shown in \cite{Had17,Rav18}. Therefore, OTFS is ideally suited to high-mobility scenarios, which encompass LEO satellite communications. 

OTFS can be implemented using the Heisenberg transform at the transmitter along with the Wigner transform at the receiver. This structure offers the possibility to implement OTFS on the basis of OFDM. Recent works provide a treatment of OTFS based on the discrete Zak transform (DZT), as detailed in \cite{Lam22}. Compared to the overlay structure, the DZT-based implementation offers reduced complexity and facilitates the application of pulse-shaping techniques. Motivated by these advantages, the work presented in this paper adopts the DZT-based OTFS (DZT-OTFS) modulation. 

The main contributions of this article are summarized as follows.
\begin{itemize}
    \item We propose a novel random access signal design that modulates the preamble into the DZT-OTFS waveform. The preamble generation method is based on concatenating identical ZC sequences. This design allows efficient use of the resources by using a single ZC root index, achieving a high level of commonality with 5G NR. A remarkable property of the proposed format is that it exhibits circularity at the receiver, even in the presence of large time offsets, without requiring long CP blocks. In fact, the CP could be entirely omitted or minimized to accommodate the maximum expected delay spread.  
\\
    \item We derive a DD domain input-output relation for the proposed random access scheme. This relation is used to characterize the impairments that affect the received preamble. To facilitate the representation of delays exceeding the length of a single ZC sequence, a dual system has been introduced.  
\\
    \item We design a preamble detection scheme tailored to the DZT-OTFS random access signal and thus operating on the DD domain. Unlike conventional two-step detection methods, the proposed scheme is able to jointly estimate fractional and integer delay components using a fixed detection window. This joint estimation approach does not require additional search steps.  Furthermore, the algorithm exploits the structure of the received signal to coherently combine the received sequences, which maximizes the desired signal strength. 
\\
    \item The numerical results demonstrate that the random access scheme presented in this paper achieves performance comparable to state-of-the-art solutions based on OFDM. Particularly, in terms of peak-to-average power ratio (PAPR) and MDP in the presence of user positioning errors. More importantly, the proposed preamble format outperforms OFDM in key aspects such as spectral confinement and overhead transmission, making it well-suited for GNSS-independent random access in NTN environments. 
\end{itemize}

\subsection{Organization}
The remainder of the paper is organized as follows. Section \ref{secII} defines the architecture, introduces the system model and characterizes the impairments of the satellite link. Then, Section \ref{secIII} describes the DZT-OTFS random access signal design and formulates the DD input-output relation. In Section \ref{secIV}, we introduce a preamble detector design that operates in the DD domain. Numerical results are shown in Section \ref{secV} and finally, the conclusions are drawn in Section \ref{secVI}.

\begin{figure}
    \centering
    \includegraphics[width=9cm]{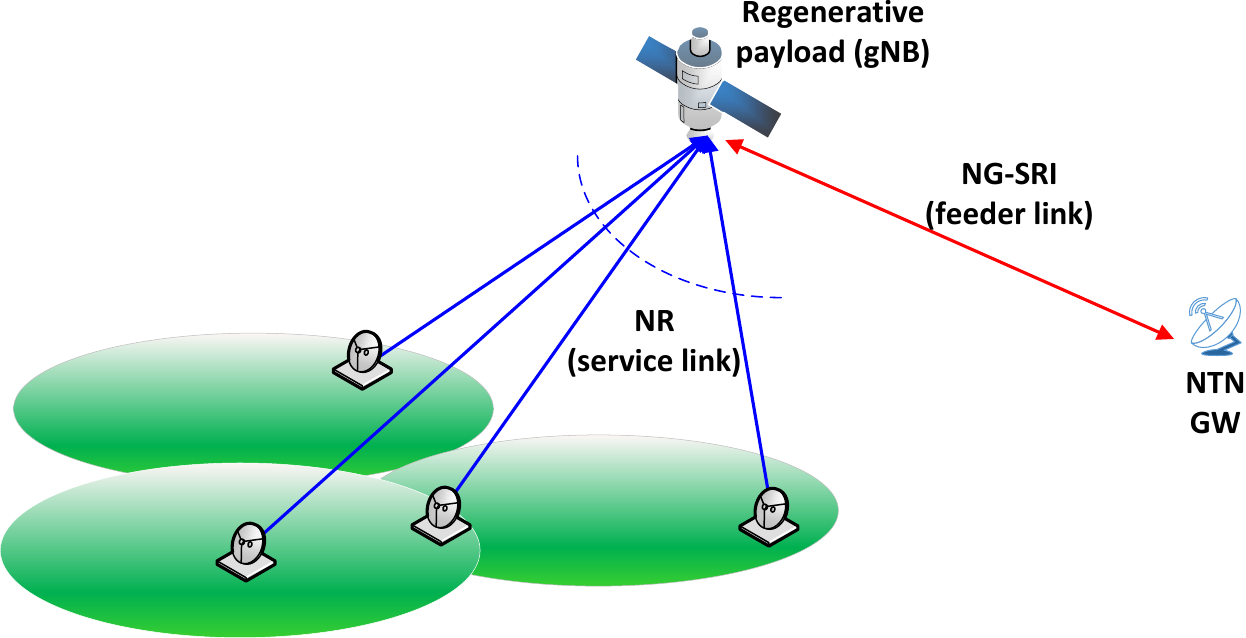}
    \caption{Satellite regenerative architecture.}
\label{fig1}
\end{figure}

\section{System model}
\label{secII}
This section is devoted to define the architecture and to formulate the system model in the uplink (UL) transmission. As illustrated in Figure \ref{fig1}, the service is provisioned by LEO satellites that divide the coverage area into multiple beams. In the proposed architecture, the satellites are integrated into the random access network (RAN). More precisely, we focus on satellites that are equipped with payloads capable of hosting a full gNB operating in the Ka band. Therefore, direct satellite connectivity is enabled without relying on terrestrial infrastructure.

\begin{figure}
\centering\includegraphics[width=8.5cm]{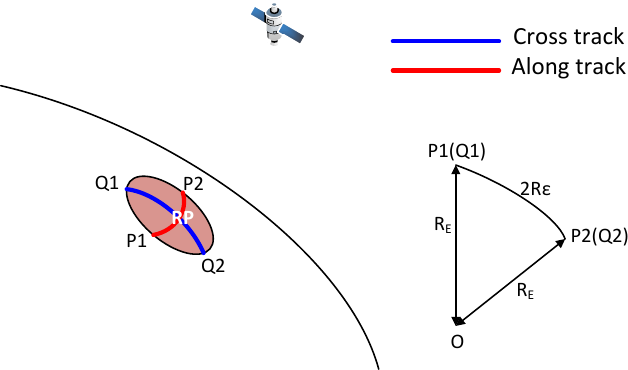}
    \caption{Geographic location uncertainty}
    \label{fig2}
\end{figure}

To transition from the idle to the connected state, the UE should perform the cell search and initiate the RAP. Upon achieving downlink synchronization and acquiring the cell identity, the UE selects a random access occasion to transmit a preamble signal, which constitutes the first step of the RAP. To ensure that the UE is received in the time and frequency resources dedicated to the physical random access channel (PRACH), the UE adjusts the timing and the carrier frequency according to the round trip time (RTT) delay and the UL Doppler frequency shift experienced in the satellite link. This pre-compensation mechanism takes as inputs the location of the UE and a valid version of the satellite ephemeris, which is acquired from the SIB19. In the presence of user positioning errors, the transmitted preamble will be time and frequency misaligned at reception. The resulting TO and CFO depend on the accuracy of the positioning information. In this work, the working assumption is that the true coordinates lie within a circular uncertainty region of radius $R_\epsilon$. The concept is illustrated in Figure \ref{fig2}, where the reference point (RP) corresponds to the true coordinates. The operating frequency, the orbit and the user positioning error will dictate the magnitude of the TO and the CFO. It shall be noticed that any displacement at high elevation angles leads to high CFO and low TO. At low elevation angles, the offsets follow an opposite trend, so that the position error translates into high TO and low CFO values. 

\subsection{Transmitter}
\begin{figure*}
    \centering
    \includegraphics[width=14cm]{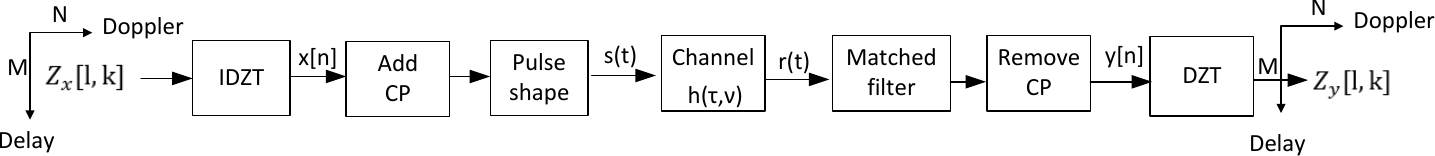}
    \caption{DZT-OTFS transmitter and receiver block diagram.}
\label{otfs}
\end{figure*}

During the attachment procedure, the UE performs the PRACH processing. A schematic representation of the transmitter and the receiver block diagram is depicted in Figure \ref{otfs}. In the first step, the UE maps $MN$ preamble symbols into the DD plane, forming the matrix $Z_{x}[l,k]$. Here $l=0,\cdots,M-1$ and $k=0,\cdots,N-1$ denote the delay and the Doppler indices, respectively. The resolutions are given by $\frac{1}{M\Delta_f}$ for delay and $\frac{1}{NT}$ for Doppler. In this work, we restrict the analysis to the case of critical sampling, i.e., $\Delta_f T=1$. The DD domain symbols are transformed into the time domain by applying the inverse DZT (IDZT), yielding the sequence  
\begin{equation}
x[l+Mm]=\frac{1}{\sqrt{N}}\sum^{N-1}_{k=0}Z_x[l,k]e^{j\frac{2\pi}{N}km},
\label{eq1}
\end{equation}
for $l=0,\cdots,M-1$ and $m=0,\cdots,N-1$. To achieve a circular structure and avoid interference from previous slots, a CP of length $L_{\text{CP}}$ is appended at the beginning of the discrete-time sequence. At the output of the CP block, the sequence can be expressed as 
\begin{equation}
x_{c}[n]=\left\{\begin{array}{cc}
 x\left[\left(n\right)_{MN} \right] & n=-L_{\text{CP}},\cdots,MN-1\\
0 & \text{otherwise}.
\end{array}\right.
\label{eq2}\end{equation}
The operation of $a$ modulo $n$ is represented by $(a)_{n}$. Then, the sequence $x_{c}[n]$ is shaped by the Nyquist filter $p(t)$, yielding a continuous time signal defined by 
\begin{equation}
s(t)=\sum^{MN+L_{\text{CP}}-1}_{n=0}x_{c}\left[n-L_{\text{CP}}\right]p\left(t-n\frac{T}{M}\right).    
\end{equation}
The transmit pulse is defined for the symbol interval $\frac{T}{M}$ and spans a time duration of $2Q\frac{T}{M}$. The interval over which the pulse has support is defined such that $2Q\ll M$. 


\subsection{Satellite channel model}
\label{secIIb}
This section models the impairments of the LEO satellite link. It must be underlined that in high-frequency bands, where directional antennas are needed, e.g., the Ka band, the multipath effect is usually weak. In such a case, the satellite channel is dominated by the line-of-sight (LoS) component. Thus, in the scenario under study, environmental effects are not
the primary source of signal degradation. Rather, the origin is the erroneous time and frequency compensation that results from the user position uncertainties. Bearing this in mind, the DD channel response is modeled as
\begin{equation}
h(\tau,\nu)=h_0\delta\left(\tau-\tau_{0}\right) \delta \left(\nu - \nu_{0} \right).   
\label{eq4}\end{equation}
The link is characterized by the channel coefficient $h_{0}$, the delay $\tau_0$, and the Doppler shift $\nu_0$. It is important to remark that the offsets $\tau_0, \nu_0$ are directly induced by inaccuracies in the UE positioning. Adopting the model proposed in \cite{Cau23}, the channel coefficient is formulated as 
\begin{equation}
h_0=\sqrt{\frac{G_{\text{UE}} G_{\text{SAT}}}{P_L K_B T_{\text{sys}} B}} e^{j\frac{2\pi}{\lambda}d_{\text{SAT-UE}}}.
\label{eq6}\end{equation}
The magnitude of the channel depends on typical gains and losses, i.e., the satellite antenna gain $G_{\text{SAT}}$, the UE antenna gain $G_{\text{UE}}$, the path loss $P_L$, the Boltzmann constant $K_B$, the bandwidth $B$ and the system noise temperature $T_{\text{sys}}$. The satellite antenna gain is assumed to be constant over the uncertainty region. The phase term depends on the carrier wavelength $\lambda$ and the slant range $d_{\text{SAT-UE}}$. 

The time and the frequency misalignment are represented as a function of the DD resolutions as 
\begin{equation}\begin{array}{cc}
\displaystyle\tau_{0}=\frac{a_0+\alpha_0}{\Delta_f M} \hspace{2em}
\displaystyle \nu_{0}=\frac{k_{0}+\kappa_{0}}{T N},
\end{array}\label{eq5}\end{equation}
where $a_{0},k_{0} \in \mathbb{Z}$ represent the integer components and $\alpha_0,\kappa_{0} \in   (-0.5, 0.5]$ are the fractional shifts.  

\subsection{Receiver}
According to the model formulated in Section \ref{secIIb}, the received signal is given by
\begin{equation}\begin{array}{rl}
r(t)&=\displaystyle\int_{\tau}\int_{\nu} h(\tau,\nu)s(t-\tau) e^{j2\pi \nu (t-\tau)} d\tau d\nu+w(t)\\
&=h_0s(t-\tau_0)e^{j2\pi \nu_0 (t-\tau_0)}+w(t).   
\label{eq7}\end{array}\end{equation}
Since the channel is normalized to the noise variance, the additive noise $w(t)$ follows a white Gaussian stochastic model with unit power spectral density. 
As shown in Figure \ref{otfs}, the received signal is passed through the matched filter, whose output is given by
\begin{equation}
y(t)=r(t)\star p^*(-t)=\int_{\tau}r(\tau)p^{*}(\tau-t)d\tau .
\label{eq8}\end{equation}
Sampling the output every $\frac{T}{M}$ seconds, we get $y\left(\left(n+L_{\text{CP}} \right)\frac{T}{M} \right)=y[n]$, for $n=0,\cdots,MN-1$. Notice that the received signal is sampled with an offset of $L_{\text{CP}}\frac{T}{M}$ to discard the CP.  

Assuming that the communication bandwidth is much larger than the maximum Doppler frequency shift, i.e., $M\Delta_f \gg v_0$, the following approximation holds
\begin{equation}\begin{array}{c}
\displaystyle \int_{\tau} p\left(\tau-\tau_0-n\frac{T}{M}\right)   e^{j2\pi \nu_0 (\tau-\tau_0)} p^{*}(\tau-t)d\tau \approx \\
\displaystyle e^{j2\pi \nu_0 n\frac{T}{M} } \int_{\tau} p\left(\tau-\tau_0-n\frac{T}{M}\right)   p^{*}(\tau-t)d\tau.
\end{array}\end{equation}
After further derivations, the received sequence becomes
\begin{equation}\begin{array}{rl}
 y[n] = &  \displaystyle \sum^{MN+L_{\text{CP}}-1}_{m=0}  h_0 R_p\left((n+L_{\text{CP}}-m)\frac{T}{M}-\tau_0\right)\times \\
 & x_{c}\left[m-L_{\text{CP}}\right] e^{j2\pi\nu_0 m \frac{T}{M}} +w[n],
\end{array}\label{eq9}\end{equation}
where $R_p(t)=p(t)\star p^{*}(-t)$. Due to the Nyquist orthogonal transmission, the noise samples are independent and identically distributed (i.i.d.) random Gaussian variables, i.e., $w[n]\sim \mathcal{CN}(0,1)$.

Since the energy of $R_p(t)$ is concentrated around its main beam, only $2L+1$ terms significantly contribute to interference, with $2L+1\ll MN$. Based on this premise and using the representation in (\ref{eq5}), we can simplify (\ref{eq9}) as 
\begin{equation}\begin{array}{rl}
 y[n] = &  \displaystyle \sum^{L}_{i=-L}  \beta[i]
  x_{c}[n-i-a_0] e^{j\frac{2\pi(k_0+\kappa_0)}{MN} \left( n-i-a_0\right)} \\
  & +w[n].
\end{array}\label{eq10}\end{equation}
To ease the analytical tractability, we define the coefficients that result from the fractionally spaced sampling by 
\begin{equation}
\beta[i]=h_0 \times e^{j 2\pi \nu_0  L_{\text{CP}}\frac{T}{M}}\times R_p\left((i-\alpha_0)\frac{T}{M}\right).
\label{eq11}\end{equation}

\subsubsection{Integer delay}
When the delay is an integer multiple of the sampling period ($\alpha_0=0$), CP transmission results in a circular structure, and (\ref{eq10}) simplifies to
\begin{equation}
 y[n] = \beta[0]
  x\left[(n-a_0)_{MN}\right] e^{j\frac{2\pi(k_0+\kappa_0)}{MN} \left( n-a_0\right)}+w[n].
\label{eq10b}\end{equation}

\subsubsection{Fractional delay}
If fractional delays are present ($\alpha_0\neq 0$), interference arises from negative and positive indices, i.e., $-L\leq i \leq L$. In this case, (\ref{eq10}) is approximated to
\begin{equation}\begin{array}{rl}
 y[n] \approx &  \displaystyle \sum^{L}_{i=-L}  \beta[i]
  x\left[(n-i-a_0)_{MN}\right] e^{j\frac{2\pi(k_0+\kappa_0)}{MN} \left( n-i-a_0\right)}\\
  & +w[n].
\end{array}\label{eq10c}\end{equation}
The equality holds true if the input sequence is defined as
\begin{equation}
x_{c}[n]=\left\{\begin{array}{cc}
 x\left[\left(n\right)_{MN} \right] & n=-L_{\text{CP}},\cdots,MN+L_{\text{CS}}-1\\
0 & \text{otherwise}.
\end{array}\right.
\end{equation}
Note that, the received signal will exhibit circularity only if a cyclic suffix (CS) of length $L_{\text{CS}}$ is appended to the transmitted sequence. However, to minimize overhead, CS is omitted in this work, and (\ref{eq10c}) is treated as an approximation.

Despite this, when the transmitted sequence conforms to (\ref{eq2}), the discrepancy between the input-output relations in (\ref{eq10}) and (\ref{eq10c}) is minimal, typically affecting only a few terms when  $i+a_0<0$. Relying on this observation, (\ref{eq10c}) is adopted, disregarding the minor modeling error. As shown in \cite{Lam22}, this interpretation is essential to benefit from DZT modulation and correlation properties, which significantly simplify the signal representation in the DD domain. From here onwards, unless otherwise stated, (\ref{eq10c}) will be regarded as the true expression. 

Regardless of the presence of fractional delays, the next step at the receiver is to apply the DZT to $y[n]$, yielding
\begin{equation}
Z_y[l,k]=\frac{1}{\sqrt{N}}\sum^{N-1}_{m=0}y[l+mM]e^{-j\frac{2\pi m}{N}k}, \label{eq12}\end{equation}
for $l=0,\cdots,M-1$ and $k=0,\cdots,N-1$. Considering the mathematical framework developed in \cite{Lam22}, the DD input-output relation can be expressed in the form of
\begin{equation}
Z_{y}[l,k]=\sum^{L}_{i=-L}\beta[i]Z_{y_i}[l,k]+Z_w[l,k],
\label{eq13b}\end{equation}
where
\begin{equation}
Z_{y_i}[l,k]=\sum^{N-1}_{m=0} \sum^{M-1}_{q=0} Z_x[q,m]Z^\tau_{i+a_0}[l-q,k] 
 Z^\nu_{k_0+\kappa_0}[q,k-m].
\label{eq13}\end{equation}
It is important to emphasize that the model considers the most general case, where the channel introduces fractional delays. Analogously to (\ref{eq12}), $Z_w[l,k]$ denotes the DZT of the noise sequence $w[n]$. Since the DZT is a unitary transform, the statistical information of the noise is kept unchanged in the DD domain. Thus, the noise samples are independent and identically distributed as follows, $Z_w[l,k]\sim \mathcal{CN}(0,1)$. To complete the model, the DD spreading functions are formulated as
\begin{equation}
Z^\tau_{a}[l,k]=\frac{1}{\sqrt{N}}\sum^{N-1}_{m=0}\delta[l-a+Mm]e^{-j\frac{2\pi}{N}mk}
\label{eq14}\end{equation}
\begin{equation}
Z^\nu_{b}[l,k]= \frac{1}{\sqrt{N}}\displaystyle e^{j\frac{2\pi}{MN}b l} e^{j\frac{\pi}{N}(N-1)(b-k)}  \displaystyle \frac{\sin\left(\pi (b-k) \right)}{\sin\left(\frac{\pi}{N} (b-k) \right)}.      
\label{eq15}\end{equation}
An important aspect to highlight is that OTFS imposes constraints on time and frequency offsets to ensure reliable detection. As detailed in \cite{Sin20}, the time and frequency misalignments are generally assumed to be bounded by $\tau_0<T$ and $-\Delta_f/2 < \nu_0 < \Delta_f/2$. In the proposed model, these constraints translate into $a_0<M$ and $-\frac{1}{2}< \frac{k_0+\kappa_0}{N}< \frac{1}{2}$. In the next section, we propose a new preamble structure specifically designed to handle scenarios with larger delays, that is, when $a_0>M$.

\begin{figure*}
    \centering
    \includegraphics[width=14cm]{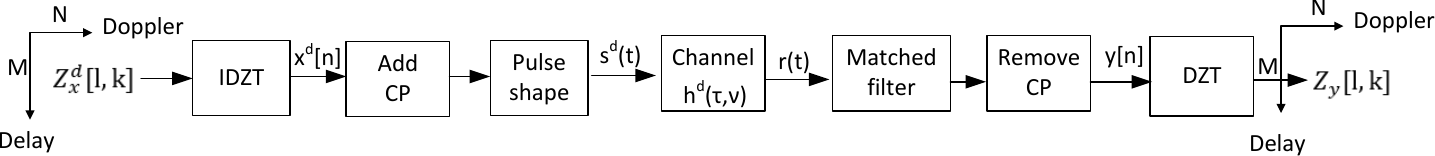}
    \caption{Dual DZT-OTFS transmitter and receiver block diagram.}
\label{otfs_dual}
\end{figure*}

\section{Preamble design}
\label{secIII}
Building upon the system model introduced in Section \ref{secII}, this section addresses the DZT-OTFS random access signal design. The objective is to enable reliable detection in the presence of large delays. Following similar design principles to those in 5G NR, the proposed scheme concatenates identical ZC sequences. However, unlike the standard, the preamble symbols are modulated in the DD domain. Hence, the OTFS frame is defined by
\begin{equation}
Z_x[l,k]=x_u [l]=e^{-j\frac{\pi u l (l+1)}{M} },    
\label{eq17}\end{equation}
for $0\leq k\leq N-1$. Notice that a ZC sequence of length $M$ is mapped into the delay domain and is repeated along the Doppler domain. The root index $u$ is selected from the set $\left\{1,\cdots,M-1\right\}$. To characterize the time domain signal, it is useful to recall that $\sum^{N-1}_{n=0} e^{j\frac{2\pi}{N}ni}=0$, for $i\in \mathbb{Z}^{+}$ and $N\geq 2$. Invoking this property, the sequence that is obtained after executing the IDZT processing becomes 
\begin{equation}
x[n]=\left\{ \begin{array}{cc}
\sqrt{N}e^{-j\frac{\pi u n (n+1)}{M} } & n=0,\cdots,M-1\\
0 & \text{otherwise}.
\end{array}\right.\label{eq18}\end{equation}
This structure offers enhanced energy efficiency by transforming the CP into a zero-padding (ZP) operation. Another remarkable property is that $x[n-a]=x\left[\left(n-a\right)_{MN}\right]$, for $0\leq a < M(N-1)$ and $0\leq n \leq MN-1$. To take advantage of this circularity at the receiver, the channel delay $\tau_0$ shall not exceed $(N-1)T$. When this constraint is satisfied, the CP block in (\ref{eq2}) can be minimized or omitted entirely, since its main purpose is to prevent the leakage from already synchronized signals that are transmitted in previous slots. Hence, the duration of the CP block should be just dimensioned only to cover the maximum expected channel delay spread, without accounting for user positioning uncertainties or additional propagation delays. Under this assumption, the normal CP length $L_{\text{CP}}$ defined in 5G NR can be used as a reference. Consequently, the total burst duration becomes $\left(L_{\text{CP}}+2Q+NM\right)\frac{T}{M}$. 

Interestingly, the study in \cite{Maz25} proposes modifications to reduce the normal CP durations in NTN scenarios. This work is motivated by the low-frequency selectivity typically observed in satellite communication channels. The analysis highlights that these adaptations could significantly reduce the overhead in future standard releases. 

At the other end of the link, the global response is defined by (\ref{eq13b}). In this section, we will show that the expression can be generalized for $0\leq a_0 < (N-1)M$, provided that the sequences are arranged in the DD domain according to (\ref{eq17}). The first step is to express the integer delay offsets as $a_0=q_M \times M+r_M$, where the quotient and the remainder are, respectively, given by $q_M=\lfloor\frac{a_0}{M} \rfloor$ and $r_M=\left(a_0\right)_M$, with $0\leq q_M <N-1$ and $0\leq r_M <M$. The second step is to consider the equivalence of the circular convolution in the delay domain. 

Closely analyzing (\ref{eq17}) and (\ref{eq18}), it can be verified that 
\begin{equation}
x\left[(n-q_M\times M)_{MN}\right] \Longleftrightarrow Z_x[l,k]e^{-j\frac{2\pi}{N}k q_M}.
\label{eq19}\end{equation}
In notation terms, (\ref{eq19}) denotes that the signal on the left (right) is the IDZT (DZT) of the signal on the right (left). 

Using this insight, we define a dual system that maintains the same original input-output relation, while simplifying the treatment of delay offsets. The corresponding block diagram is illustrated in Figure \ref{otfs_dual}. The difference lies in the treatment of timing errors. Under this alternative view, the dual channel response is represented by 
\begin{equation}
h^d(\tau,\nu)=h^d_0\delta\left(\tau-\tau^d_{0}\right) \delta \left(\nu - \nu^d_{0} \right),
\end{equation}
with 
\begin{equation}\begin{array}{cc}
h^d_0=h_0 e^{-j\frac{2\pi(k_0+\kappa_0)}{N}q_M} \hspace{2em} \displaystyle\tau^d_{0}=\frac{r_M+\alpha_0}{\Delta_f M} \hspace{2em}
\displaystyle \nu^d_{0}=\nu_{0}.
\end{array}\label{eq5b}\end{equation}
Here, the offset $q_M\times M$ is not induced by the channel, but it is instead deliberately embedded at the transmitter. Accordingly, the transmitted sequence in the dual system is given by 
\begin{equation}
x^d[l+Mm]=\frac{1}{\sqrt{N}}\sum^{N-1}_{k=0}Z^d_x[l,k]e^{j\frac{2\pi}{N}km},
\label{eq20}
\end{equation}
where 
\begin{equation}
Z^d_x[l,k]=Z_x[l,k]e^{-j\frac{2\pi}{N}k q_M}.
\label{eq19b}\end{equation}
After transmission through the channel defined by $h^d(\tau,\nu)$, the received signal can be expressed as
\begin{equation}\begin{array}{rl}
 y[n] = & \displaystyle \sum^{L}_{i=-L}  \beta[i]
  x^d\left[(n-i-r_M)_{MN}\right] e^{-j\frac{2\pi(k_0+\kappa_0)}{N}q_M} \times \\
 & e^{j\frac{2\pi(k_0+\kappa_0)}{MN} \left( n-i-r_M\right)}
  +w[n].
\end{array}\label{eq21}\end{equation}
It can be verified that (\ref{eq10c}) and (\ref{eq21}) are equivalent, as $x\left[(n-i-a_0)_{MN}\right]=x^d\left[(n-i-r_M)_{MN}\right]$. For notational simplicity, we adopt the dual system representation. Under this formulation, the DD domain output is given by 
\begin{equation}
Z_{y}[l,k]=e^{-j\frac{2\pi(k_0+\kappa_0)}{N}q_M}\sum^{L}_{i=-L}\beta[i] Z_{y_i}[l,k]+Z_w[l,k],
\label{eq22}\end{equation}
where the DD input-output relation for each multipath tap is expressed as
\begin{equation}
Z_{y_i}[l,k]=\sum^{N-1}_{m=0} \sum^{M-1}_{q=0} Z^d_x[q,m]Z^\tau_{i+r_M}[l-q,k] 
 Z^\nu_{k_0+\kappa_0}[q,k-m].
\label{eq23}\end{equation}
The new model, in conjunction with (\ref{eq14}) and (\ref{eq15}), is employed to derive (\ref{eq16}). To simplify the notation, it is assumed that $0\leq i+r_M \leq M-1$. In the next section, we will demonstrate that the proposed detection algorithm can be effectively developed based on the formulation in (\ref{eq16}).

\begin{figure*}
\begin{equation}
Z_{y_i}[l,k]=\left\{\begin{array}{rl}
\displaystyle \frac{e^{-j\frac{2\pi}{N}k}}{\sqrt{N}}\sum^{N-1}_{m=0}  Z^d_x[\left(l-i-r_M\right)_M,m]Z^\nu_{k_0+\kappa_0}[\left(l-i-r_M\right)_M,k-m] & \text{if} \ \ 0\leq l<i+r_M\\
\displaystyle \frac{1}{\sqrt{N}}\sum^{N-1}_{m=0} Z^d_x[\left(l-i-r_M\right)_M,m]Z^\nu_{k_0+\kappa_0}[\left(l-i-r_M\right])_M,k-m] & \text{if} \ \ i+r_M\leq l \leq  M-1
\end{array}\right.\label{eq16}\end{equation}
\hrulefill
\vspace{-5mm}
\end{figure*}
\section{Preamble detection}
\label{secIV}
The objective of this section is to design a preamble detector specifically tailored to the format introduced in Section \ref{secIII}. It will be shown that the delays experienced by each transmission can be easily inferred from the detected preambles. To design the detection algorithm, we build upon the model described in Section \ref{secIII}. In this regard, the detector takes as input the signal $Z_y[l,k]$, defined in (\ref{eq22}), which reveals that the DD domain output is affected by inter-delay and inter-Doppler interference. The key insight is summarized in the following proposition. 

\begin{pro}
If the ZC sequence is arranged in the DD domain according to (\ref{eq19b}), then the interfering terms in the Doppler domain add constructively. \label{prop1}\end{pro}

The mathematical details that lead to the proof of Proposition \ref{prop1} are provided hereinafter. For a given delay index $l$ and a Doppler shift $\nu_0=b/NT$, the following equivalence can be established:
\begin{equation}\begin{array}{l}
\displaystyle \sum^{N-1}_{m=0}  \frac{1}{\sqrt{N}}Z^d_x[l,m]Z^\nu_{b}[l,k-m]\\
\displaystyle =\frac{x_u[l]e^{j\frac{2\pi}{MN}b l}}{N}
\displaystyle \sum^{N-1}_{m=0} e^{-j\frac{2\pi}{N}mq_M} \sum^{N-1}_{i=0} e^{j\frac{2\pi}{N}(b-k+m)i}\\
\displaystyle =x_u [l] e^{j\frac{2\pi}{MN}b l}\displaystyle  e^{j\frac{2\pi}{N}(b-k)q_M}.
\end{array}\label{eq24}\end{equation}
An immediate consequence of Proposition \ref{prop1} is that the summation in (\ref{eq16}) can be eliminated. Applying this result, we can define a new system equation, namely,
\begin{equation}\begin{array}{rl}
Z_{y}[l,k]=&\displaystyle \sum^{L}_{i=-L}\beta[i] x_{uk} \left[ \left(l-i-r_M\right)_{2M}\right] \times \\
& e^{j\frac{2\pi}{MN}(k_0+\kappa_0) l}  e^{-j\frac{2\pi}{N}kq_M}+Z_w[l,k].
\end{array}\label{eq25}\end{equation}
To characterize the phase rotation in $l=0,\cdots,i+r_M$, we establish a periodization of the extended signal
\begin{equation}
x_{uk}[l]=\left\{\begin{array}{cc}  
x_u[l] & 0\leq l \leq M-1\\ 
e^{-j\frac{2\pi}{N}k}x_u[l] & M \leq l \leq 2M-1,
\end{array}\right. \label{eq26}\end{equation}
which is made up of two concatenated ZC sequences. The key parameters of (\ref{eq25}) are $q_M$ and $r_M$, which have been defined to express the total delay as $a_0=q_M\times M+r_M$. Hence, the objective of the detector is to find the signature of the transmitted preamble by jointly searching over $0\leq q_M < N-1$ and $0\leq r_M \leq M-1$. Upon detecting the presence of the preamble, the detection algorithm shall return a two-dimensional index. Next, the detector will provide an estimate of the delay, which can be unambiguously derived from the selected index. 

The analysis is initially restricted to scenarios where the offsets span this range $0\leq i+r_M \leq M-1$. The rest of the cases where the indices may not be confined within that interval will be examined and addressed in subsequent analyses. 

Assuming the working hypothesis $0\leq i+r_M \leq M-1$, the decision variable for preamble detection is computed as
\begin{equation}\begin{array}{rl}
    \rho_{\upsilon}(\mu,\gamma)=&\displaystyle \left|\sum^{N-1}_{k=0}\sum^{M-1}_{l=0}e^{j\frac{2\pi}{N}k \gamma}\frac{Z_{y}[l,k]}{M N}x^{*}_{\upsilon k}\left[(l-\mu)_{2M} \right] \right|^2 \\
    =& \displaystyle \frac{1}{|M N|^2}\left|C_{\upsilon }(\mu,\gamma)\right|^2+\eta_{\upsilon}(\mu,\gamma),
\end{array}\label{eq27}\end{equation}
where the correlation term is
\begin{equation}\begin{array}{c}
C_{\upsilon }(\mu,\gamma)=  \displaystyle \sum^{N-1}_{k=0}\sum^{L}_{i=-L}\sum^{M-1}_{l=0}\beta[i]e^{j\frac{2\pi}{N}k(\gamma-q_M)}\times  \\
 x^*_{\upsilon k}\left[\left(l-\mu\right)_{2M}\right] x_{uk}\left[\left(l-i-r_M\right)_{2M}\right] e^{j\frac{2\pi (k_0+\kappa_0)}{MN} l}.    
\end{array}\label{eq28}\end{equation}
The noise component in (\ref{eq27}) is defined by $\eta_{\upsilon}(\mu,\gamma)$, where the variable $\upsilon$ denotes the root index of the candidate ZC sequence. The standard approach is to use 64 ZC sequences to generate the preamble set. To account for all possible delay and Doppler shifts, the decision variable in (\ref{eq27}) is evaluated over the ranges $0\leq \mu \leq M-1$ and $0\leq \gamma < N-1$. Since the identical ZC sequence is transmitted across the Doppler domain, the decision variable results from the coherent accumulation process. It is important to remark that, in OFDM systems, coherent combination of sequences is not possible in the presence of CFO, due to the accumulated phase rotation across each multicarrier symbol. Consequently, conventional
detection methods for repeated sequences in OFDM rely on non-coherent accumulation, as shown in \cite{Cau24b,Ses11}. Thanks to the specific characteristics of OTFS, the detector devised in this section can coherently combine the received signals across
the Doppler domain, enabling more effective computation of the decision variable. 

\subsection{Practical implementation}
\begin{figure}
    \centering
    \includegraphics[width=9cm]{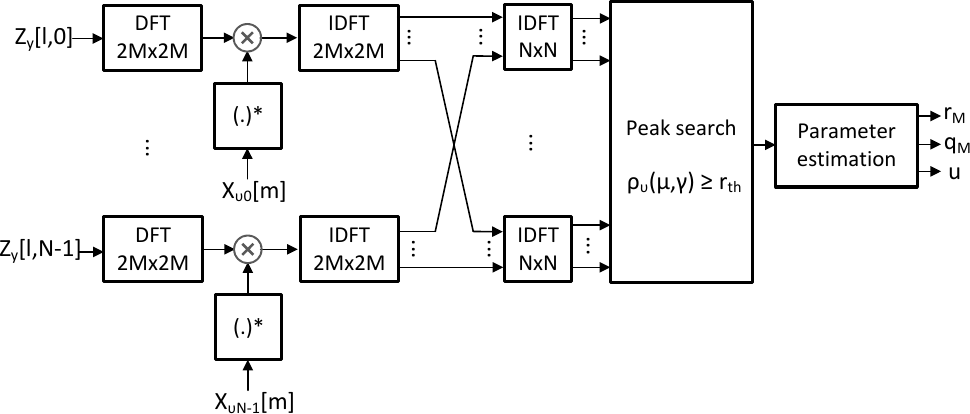}
    \caption{Detection scheme.}
\label{fig4}
\end{figure}

It is noteworthy to mention that the variable defined in (\ref{eq27}) is the result of combining $N$ circular convolutions. Therefore, $\rho_{\upsilon}(\mu,\gamma)$ can be efficiently implemented using IDFT and DFT blocks. To reduce the computational complexity, the sizes of DFT and IDFT blocks can be selected as a power of 2. The schematic view of the detector is shown in Figure \ref{fig4}. The whole processing is divided into four steps. The overall computational complexity can be analyzed by examining each step. First of all, the columns of $Z_{y}[l,k]$ are transformed by means of a DFT of size $2M$. This results in a complexity of $\mathcal{O}(VNM \log M)$, where $V$ is the number of candidate ZC sequences. The next step is to compute the pointwise multiplication with $X^*_{\upsilon k}[m]$, which represents the DFT of $x^*_{\upsilon k}[-l]$. This operation has a linear complexity of $\mathcal{O}(VNM)$. Then, the resulting signal is transformed back to the delay domain by executing an IDFT of 2M points. Bearing in mind the range of $\mu$, we just need to take the first $M$ IDFT outputs, contributing with a complexity of $\mathcal{O}(VNM \log M)$. Finally, the signals processed in different Doppler bins are fed into an N-point IDFT to compute $\rho_{\upsilon}(\mu, \gamma)$, adding $\mathcal{O}(VMN \log N)$ to the complexity. Hence, the overall computational complexity is given by $\mathcal{O}\left(VNM \log(MN)\right)$.

In the detection scheme, the time resolution is $\frac{1}{M\Delta_f}$, which is controlled by the 2M-point IDFT. Hence, finer time resolution can be achieved by increasing the number of points.

The two-step detection approach has complexity of order $\mathcal{O}\left(VNM \log (M) \right)$ for estimating fractional delays. This fractional value is then used to shift the DFT window and capture integral ZC sequences. Upon applying this alignment, the integer component is obtained by detecting the boundaries of the random access signal. Given that the integer delay estimation involves additional demodulation steps and correlations, the overall complexity remains comparable to that of the proposed detector.

\subsection{Analysis of the correlation}
Although the sequence defined in (\ref{eq26}) does not strictly preserve the ZC correlation properties, it is observed that the magnitude of (\ref{eq27}) remains small for $\upsilon\neq u$. For the case $\upsilon=u$, a more thorough analysis is conducted to gain further insight into the correlation properties. Accordingly, we start analyzing the special case where the delay is integer and the Doppler frequency shift is zero. In such a case, the most relevant finding is stated in Proposition \ref{prop2}. 

\begin{pro}
Assuming that $\alpha_0=0$, $k_0+\kappa_0=0$ and $\upsilon=u$, the correlation defined in (\ref{eq28}) approximates an orthogonal function. \label{prop2}\end{pro}

\begin{figure}
    \centering
    \includegraphics[width=8cm]{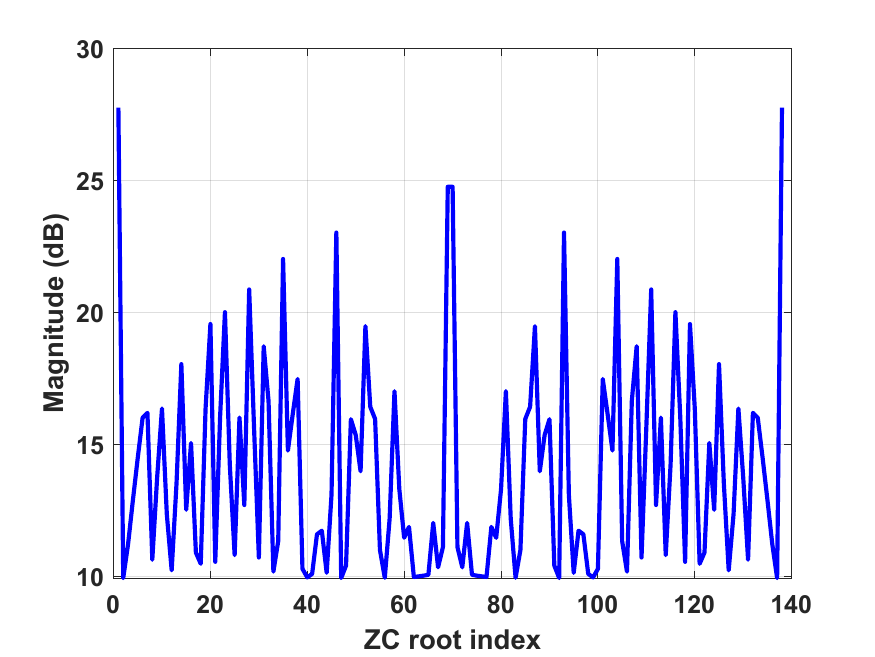}
    \caption{Power ratio between the main peak and the main pseudo-peak as function of the ZC root index, for $M=139$.}
\label{fig3}
\end{figure}

To ascertain that the approximation is reasonably accurate, we particularize the correlation to $\alpha_0=0$, $k_0+\kappa_0=0$ and $\upsilon=u$, yielding
\begin{equation}\begin{array}{rl}
C_{u}(\mu,\gamma)=&  \displaystyle \sum^{N-1}_{k=0}\sum^{M-1}_{l=0}h_0 e^{j\frac{2\pi}{N}k(\gamma-q_M)}\times  \\
& x^*_{uk}\left[\left(l-\mu\right)_{2M}\right]x_{uk}\left[\left(l-r_M\right)_{2M}\right].    
\end{array}\label{eq30}\end{equation}
To obtain the closed-form expression, we can exploit a key property of ZC sequences, whereby the cyclic shifted version of a root sequence satisfies $x_u\left((l+\mu)_M\right)\propto e^{-j\frac{2\pi}{M}u\mu l}x_u(l)$. Using this result, we get
\begin{equation}
\begin{array}{l}
|C_{u}(\mu,\gamma)|=\\
\left\{ \begin{array}{cc}
MN|h_0| & \mu=r_M, \gamma=q_M \\ 
 \displaystyle \left| \sum^{|r_M-\mu|-1}_{l=0} Nh_0 e^{j\frac{2\pi}{M}u(r_M-\mu)l}\right| & (\mu,\gamma)\in \mathbb{S}_{\mu,\gamma}\\ 
0 & \text{otherwise},
\end{array}\right. \end{array}\end{equation}
where 
\begin{equation}\begin{array}{rl}
\mathbb{S}_{\mu,\gamma}=&\displaystyle\left\{ \left\{\mu\neq r_M, \gamma=q_M\right\} \cup \left\{\mu>r_M,\gamma=(q_M-1)_N \right\} \right. \\ &\left.
 \displaystyle\cup \left\{\mu<r_M,\gamma=(q_M+1)_N \right\}\right\}.    
\end{array}\end{equation}
This result reveals that many correlation entries are zero, facilitating detection. From the non-zero entries, it can be inferred that pseudo-peaks may still emerge. In this regard, Figure \ref{fig3} shows that the power ratio between the main peak and the main pseudo-peak is always higher than 10 dB. The higher the power ratio, the better the correlation characteristics. 

In light of the above discussion, we can conclude that the approximation stated in Proposition \ref{prop2} holds true. 

Moving beyond ideal conditions, it is worth analyzing the case where the fractional delay is not equal to zero, i.e., $\alpha_0\neq 0$. In such event, we can infer from Proposition \ref{prop2} that correlation peaks emerge at $\mu=\left\{ -L+r_M,\cdots,L+r_M\right\}$ and $\gamma=q_M$. Since the magnitude of $\beta[i]$ will reach the maximum value at $i=0$, it follows that the highest correlation peak will be observed at $\mu=r_M$. In this specific case, i.e., when $i=0$, the condition $0\leq i+r_M \leq M-1$ is always satisfied. Hence, it is not necessary to address the special cases where $i+r_M<0$ or $i+r_M>M-1$. It deserves to be mentioned that in the most detrimental situation where $\alpha=\pm 0.5$, the magnitude of $\beta[0]$ takes the lowest possible value, which depends on the pulse autocorrelation function. Another issue worth considering is the small power imbalance between $\rho_u(r_M,q_M)$ and $\rho_u(r_M\pm 1,q_M)$, for $\alpha=\pm 0.5$. In this situation, the detector may select a peak that is adjacent to the true value. Accordingly, the wrong detection translates into a timing error that is equal to $\pm \frac{1}{M\Delta_f}$. Although the error is small, Section \ref{secIVC} discusses refinement methods to mitigate this effect. An important conclusion that can be drawn from this analysis is that the performance of the detector will be degraded in the presence of fractional delays.   

Concerning the adverse effects induced by the CFO, it becomes evident that as $k_0+\kappa_0$ increases, the power of the main peak reduces. That is because the energy will spread to all lag indices $\mu$. 
Under these conditions, the correlation at $\mu=r_M$ and $\gamma=q_M$ becomes 
\begin{equation}\begin{array}{rl}
C_{u}(r_M,q_M)=& N\beta[0] \frac{\sin\left( \frac{\pi (k_0+\kappa_0)}{N}\right)}{\sin\left( \frac{\pi (k_0+\kappa_0)}{MN}\right)}  e^{j\frac{\pi (k_0+\kappa_0) (M-1)}{MN} } \\
& \displaystyle +\sum^{N-1}_{k=0}\sum_{i\neq 0}\sum^{M-1}_{l=0}\beta[i]x^*_{u k}\left[\left(l-r_M\right)_{2M}\right]   \\
& \times x_{uk}\left[\left(l-i-r_M\right)_{2M}\right] e^{j\frac{2\pi (k_0+\kappa_0)}{MN} l}.    
\end{array}\label{eq29}\end{equation}
The combined effects of the fractional delay and the CFO may cause difficulty in detecting the peak and eventually leading to erroneous detection. To limit the performance degradation that results from the CFO, the waveform parameters shall be set so that $|\frac{k_0+\kappa_0}{N}|<1/2$. Otherwise, the main peak could be excessively attenuated, rendering the detector ineffective.  

\begin{algorithm}[t]
\SetAlgoLined

\KwIn{
\\- Received signal matrix $Z_y[l,k]$ \\
- Preamble set of $V$ ZC sequences \\
- Detection threshold $r_{\mathrm{th}}$
}
\KwOut{
Estimated parameters $(\hat{u}, \hat{r}_M, \hat{q}_M)$
}

\BlankLine
\For{$\upsilon=0:V-1$}{
\For{$k=0:N-1$}{
    \textbf{Step 1: Frequency-Domain Transformation}\;\\
    Perform $2M$-point DFT on each column of $Z_y[l,k]$ to obtain $Z_y^{\mathrm{DFT}}[m,k]$\;

    \textbf{Step 2: Pointwise Multiplication}\;\\
    \For{$m=0:2M-1$}{
        $Y_{\upsilon}[m,k]=Z_y^{\mathrm{DFT}}[m,k] \cdot X^*_{\upsilon k}[m]$\;
    }

    \textbf{Step 3: Delay-Domain Transformation}\;\\
    Perform $2M$-point pruned IDFT (second half zero) on each column of $Y_{\upsilon}[m,k]$ to obtain $y_{\upsilon}[\mu,k]$\;

    \textbf{Step 4: Coherent Accumulation}\;\\
    \For{$\mu=0:M-1$}{
        Perform $N$-point IDFT on each row of $y_{\upsilon}[\mu,k]$ and compute the absolute square to obtain $\rho_{\upsilon}(\mu,\gamma)$\;
    }
}
}

\BlankLine
\textbf{Step 5: Peak Search and Parameter Estimation}\;\\
\textbf{Initialize:} $\hat{\rho}=0$, $(\hat{u}, \hat{r}_M, \hat{q}_M)=(0, 0, 0)$\;

\For{$\upsilon=0:V-1$}{ 
    \For{$\mu=0:M-1$}{ 
        \For{$\gamma=0:N-1$}{ 
            \If{$\rho_{\upsilon}(\mu,\gamma) \geq r_{\mathrm{th}}$ \textbf{and} $\rho_{\upsilon}(\mu,\gamma) > \hat{\rho}$}{
                $\hat{\rho}=\rho_{\upsilon}(\mu,\gamma)$,\;
                $(\hat{u}, \hat{r}_M, \hat{q}_M)=(\upsilon, \mu, \gamma)$\;
            }
        }
    }
}
\BlankLine
\textbf{Step 6: Fractional Delay Refinement}\;

\If{ $\frac{\rho_{\hat{u}}(\hat{r}_M, \hat{q}_M)}{\rho_{\hat{u}}(\hat{r}_M\pm 1, \hat{q}_M)}< 1.25$}{
    Adjust delay estimate: $\hat{r}_M=\hat{r}_M \pm 0.5$\;
}

\BlankLine
\Return $(\hat{u}, \hat{r}_M, \hat{q}_M)$\;

\caption{Detection Scheme}
\label{alg:detection}
\end{algorithm}

\subsection{Peak search and parameter estimation}
\label{secIVC}
The detector is able to characterize the signature of the transmitted preamble by solving 
\begin{equation}
\left\{\hat{u},\hat{r}_M,\hat{q}_M\right\}=\operatorname*{argmax}_{\upsilon,\mu,\gamma} \text{P}\left( \rho_{\upsilon}(\mu,\gamma) \geq r_{\text{th}}\right).    
\label{eq31}\end{equation}
The distinctive feature of this approach is that $r_M$ and $q_M$ are jointly estimated. The common approach is to separately estimate these two parameters that define the delay. Several examples can be found in \cite{Zhu24,Zhe18,Cau24,Cau24b}. The consequence of carrying out sequential processing is that errors are propagated. Hence, the estimation of $r_M$ may become the performance bottleneck. In this regard, a small deviation in $r_M$ could lead to a wrong decision in $q_M$, as shown in \cite{Cau24b}. The joint detection proposed here addresses this issue by accounting for the dependencies between the parameters. 

The criterion that drives the parameter estimation in (\ref{eq31}) is based on finding the index that exhibits the highest peak value. The magnitudes that are above the detection threshold will be classified as peak values. The parameter $r_{\text{th}}$ is computed to comply with the false alarm probability, which is defined by
\begin{equation}
P_{\text{FA}}=\text{P}\left( \eta_{\upsilon}(\mu,\gamma) \geq r_{\text{th}}\right).
\end{equation}
Bearing in mind the statistical information provided in Section \ref{secIII} and the closed-form expression (\ref{eq27}), we can resolve that the noise component can be formulated as $\eta_{\upsilon}(\mu,\gamma)=\frac{1}{MN} \chi^2(2)$. The term $\chi^2(n)$ is used to represent a chi-square distribution with $n$ degrees of freedom. The equation that relates $P_{\text{FA}}$ to $r_{\text{th}}$ is derived according to \cite{Roh16} as
\begin{equation}
P_{\text{FA}}=1-\left(1-e^{-\frac{r_{\text{th}}MN}{2}}\right)^M.
\end{equation}
It is important to note that the detection process defined by (\ref{eq31}) can be impaired by the leakage effect resulting from the fractional delay. This is clearly seen in (\ref{eq11}) for $\alpha_0=\pm 0.5$. In such cases, it is highly likely that two adjacent indices, such as $\mu^*$ (the true peak) and its neighbors $\mu^*\pm 1$, may both be declared as peaks. Without corrective measures, the detector may incorrectly identify a neighboring index as the true peak. 

Although the impact on the estimation accuracy is not significant, we propose a low complexity solution to refine the estimates. The approach is based on evaluating the power imbalance between the correlation peaks observed at $\mu^*$ and $\mu^*\pm 1$. If the difference is relatively small, e.g., less than 25 $\%$, then the outcome is the average of the peak positions. The resulting index becomes $\mu^*\pm 0.5$. The sign of the additive term depends on whether the fractional delay is positive or negative. Notice that the proposed method improves the resolution, without resorting to complex techniques such as fractionally spaced sampling receivers. A pseudo-code of the detection algorithm is described in Algorithm \ref{alg:detection}.

To sum up, the detection scheme envisaged in this section benefits from joint detection and coherent signal combining. Nevertheless, the scheme suffers from distortion in the presence of fractional delays. To understand which is the dominant effect, some numerical results are presented in the next section.

\section{Numerical results and discussion}
\label{secV}
In this section, numerical simulations are presented. We first analyze the PAPR and the frequency confinement of the transmitted signal, and then the MDP is computed. We have applied the methods described in Sections \ref{secIII} and \ref{secIV} to respectively generate and detect the random access signal. To achieve a high level of commonality with 3GPP, 64 ZC sequences are formed following the procedure specified in \cite{nr211} for FR2. 
To shape the preamble symbols, we have employed the square-root raised cosine filter with a roll-off factor of 10$\%$ and finite support in the time domain. To confine the energy of the burst within a short interval, the pulse spans $20\frac{T}{M}$, which is parametrized by setting $2Q=20$. 

\begin{table}
\centering
\caption{System and orbit parameters}
	\begin{tabular}{|l|l|l|}
	\hline
		Frame size & $N=4$, $M=139$ \\ \hline
        DD resolutions & $T=16.67 \mu s$, $\Delta_f=60$ kHz \\ \hline
        PRACH bandwidth   & $B=8.34$ MHz \\ \hline
        Frequency band & Ka \\ \hline
		Satellite altitude & 550 km \\ \hline
		Minimum elevation angle & 30º  \\ \hline
	\end{tabular}
\label{table1}
\end{table}

\begin{table}
\centering
\caption{PRACH parameters}
	\begin{tabular}{|l|c|c|c|}
	\hline
	                             & OTFS & OFDM \cite{Cau24b} & OFDM \cite{Zhu24} \\ \hline
	   Preamble length  & $\frac{T}{M}(M+2Q)$ &  $NT$  & $(N+1)T$ \\ \hline
	   CP  length & 0 & $(N-1)T$ & 0\\ \hline
	   Guard time   & $(N-1)T$ & $(N-1)T$ & $(N-1)T$\\ \hline
	   Detection window & $NT$ & $NT$ & $2NT$ \\ \hline
      Number of roots &  1 & 1 & 2 \\ \hline
	   \shortstack{Detection\\ algorithm} & \shortstack{One-step \\ CA} & \shortstack{Two-step\\ NCA} & \shortstack{Two-step\\ CA} \\ \hline
       \shortstack{Detector\\ time resolution} & $\displaystyle\frac{1}{512\Delta_f}$ & $\displaystyle\frac{1}{512\Delta_f}$ & $\displaystyle\frac{1}{512\Delta_f}$ \\ \hline
	\end{tabular}
\label{table2}
\end{table}

In the scenario under study, we focus on LEO satellite systems with regenerative capabilities. Hence, the RTT delay and the Doppler frequency shift exclude feeder link impairments. The system and orbit parameters are gathered in Table \ref{table1}. To ensure that the preamble is received within the allocated RACH resources, a pre-compensation mechanism is autonomously applied by the UE. Given the low delay spread in satellite channels, preamble arrival differences
mainly arise from UE positioning errors. Based on that, it is assumed that already synchronized signals virtually do not interfere with the received preambles. Accordingly, the CP length is set to zero. Nevertheless, to ensure reliable detection, the duration of both the preamble and the guard time shall be larger than the residual delay. Following this rationale and bearing in mind the design principles described in Section \ref{secIII}, the delay should not exceed $(N-1)T$. Accordingly, the guard time is equal to $(N-1)T$. The details are provided in Table \ref{table2}.

\begin{figure}
    \centering
    \includegraphics[width=8cm]{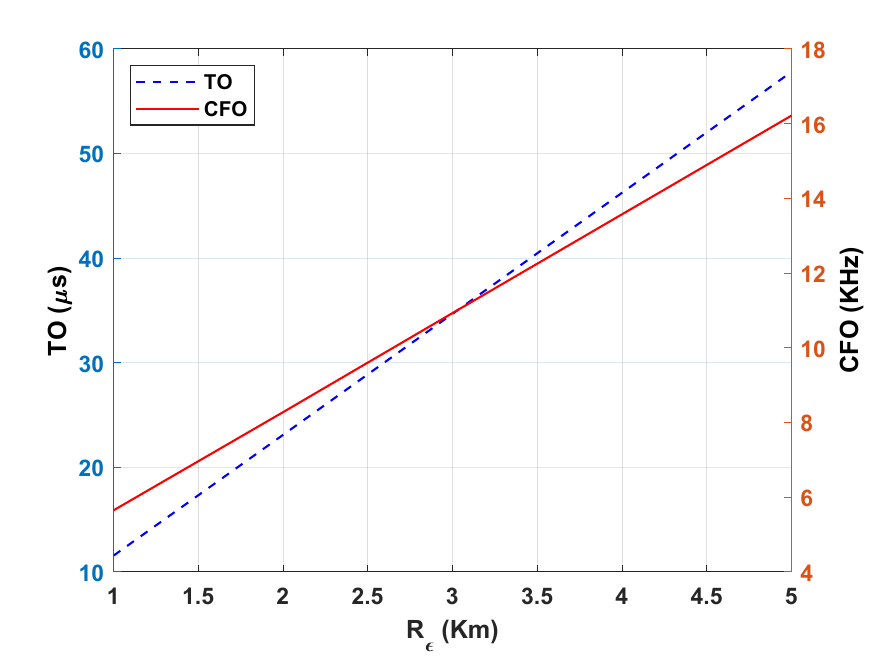}
    \caption{Maximum TO and CFO values as function of $R_\epsilon$.}
\label{fig5}
\end{figure}

Following the methodology outlined in \cite{Cau24}, we compute the maximum TO and CFO values that result from UE positioning errors. The corresponding results are represented in Figure \ref{fig5} as a function of the radius of the uncertainty region $R_\epsilon$. As discussed in Section \ref{secIIb}, timing and frequency offsets are characterized by the parameters $\tau_0$ and $\nu_0$. Under the proposed configuration, preamble detection is feasible as long as $0\leq \tau_0 <(N-1)T$ and $-\Delta_f/2 \leq \nu_0 \leq \Delta_f/2$). Evaluating these conditions for the parameters used in this section, it follows that the uncertainty region should not exceed $R_\epsilon=4.3$ km.

As a benchmark, we have considered the solutions proposed in \cite{Cau24b,Zhu24}. Both schemes modulate ZC sequences using DFT-s-OFDM. Analogously to the method described in this work, the preamble provided in \cite{Cau24b} is obtained by concatenating $N$ identical ZC sequences. The main difference stems from the fact that a CP block of length $(N-1)T$ is needed to achieve a circular structure at reception. To estimate the delay, 
the detector uses a fixed detection window of length $NT$, which is precisely the preamble duration. 
The random access signal structure is enhanced in \cite{Zhu24} by removing the CP. Interestingly, the preamble is constructed by using two ZC root sequences. The resulting preamble spans 
$(N+1)T$. At reception, the signal does not exhibit circularity. Hence, to capture the preamble samples, the detection window should cover all possible delays. Accordingly, the window length should be at least $2NT$. Concerning the detection algorithm, both schemes resort to the two-step design, where fractional and integer delay components are separately estimated. However, a clear distinction shall be made between the techniques regarding the computation of the decision variable. The solution conceived in \cite{Cau24b} performs non-coherent accumulation (NCA). By contrast, the authors in \cite{Zhu24} rely on coherent accumulation (CA). The specific characteristics of each design are described in Table \ref{table2}. From here onwards, we will refer to OFDM NCA and OFDM CA to identify the techniques described in \cite{Cau24b} and \cite{Zhu24}, respectively. Based on the values gathered in Table \ref{table2}, we can conclude that OFDM-based designs require a considerably larger amount of resources to transmit and receive the preamble than the proposed scheme.

\begin{figure}[t]
    \centering
    \includegraphics[width=0.5\textwidth]{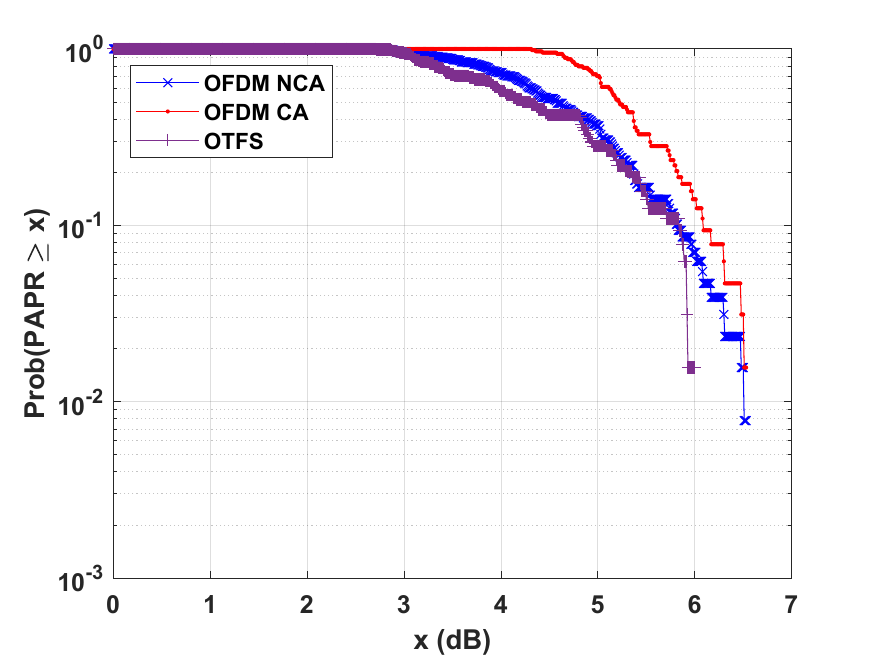}
    \caption{CCFD of PAPR of random access signals.}
    \label{fig_papr}
\end{figure}

\subsection{Peak-to-average power ratio}
This subsection focuses on evaluating the envelope fluctuation of the random access signals. To this end, we have represented in Figure \ref{fig_papr} the complementary cumulative distribution function (CCDF) of the PAPR. The sequences are generated by considering all ZC root indices. As Figure \ref{fig_papr} shows, OTFS demonstrates a slightly better PAPR performance compared to OFDM. These results indicate that, in OTFS modulation schemes, the power amplifier efficiency will not differ significantly from that of OFDM.

\begin{figure}
    \centering
    \includegraphics[width=0.5\textwidth]{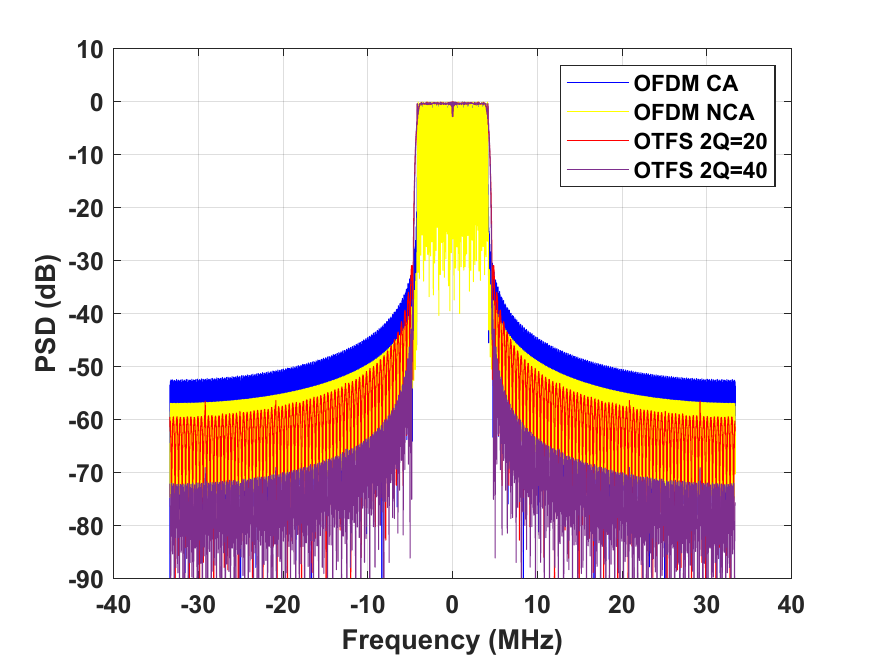}
    \caption{PSD of random access signals.}
    \label{psd}
\end{figure}

\subsection{Frequency response}
In this subsection, we aim at evaluating the spectral leakage suppression outside the PRACH bandwidth. To evaluate the frequency domain information, Figure \ref{psd} depicts the power spectral density (PSD) in a 60 MHz bandwidth. For the sake of clarity, the highest spectral peak is normalized to 0 dB. An interesting result is that the PSD decays faster in OFDM NCA than in OFDM CA. This can be attributed to the CP that is transmitted in OFDM NCA, which extends the rectangular pulse that is used to shape the subcarrier signals. Hence, by increasing the CP length, the subcarrier signals become narrower in the frequency domain, reducing out-of-band (OOB) emissions. However, the overhead is significantly increased. By contrast, thanks to pulse shaping, OTFS can improve the spectrum confinement with a moderate overhead. By increasing the duration of the prototype pulse, which is controlled by the variable $Q$, OTFS exhibits much higher stopband
attenuation than OFDM. 

\subsection{Missed detection probability}
To complete the performance analysis, we evaluate the MDP versus SNR. The transmission goes through the channel model that is defined in Section \ref{secII}. Hence, the SNR is a function of the satellite channel gain, the power of the symbols and the power of the noise, i.e., $\text{SNR}=|h_0|^2 \mathbb{E}\left\{|Z_x[l,k]|^2 \right\}$. The events that lead to an erroneous detection include: 1) the detection of a different preamble than the one that was sent, 2) not detecting a preamble at all or 3) correct preamble detection but with the wrong timing estimation. Specifically, a timing estimation error is declared if the magnitude of the error exceeds $\frac{1}{M\Delta_f}$. We consider that $N_U$ users are simultaneously accessing the network in a single random
access occasion. In alignment with the PRACH study cases defined in \cite{nr821}, we consider $N_U=1,2$. All detection schemes that are assessed in this section are configured to ensure a false alarm probability of $P_{\text{FA}}=10^{-3}$.

\begin{figure}[t]
\centering
\includegraphics[width=0.5\textwidth]{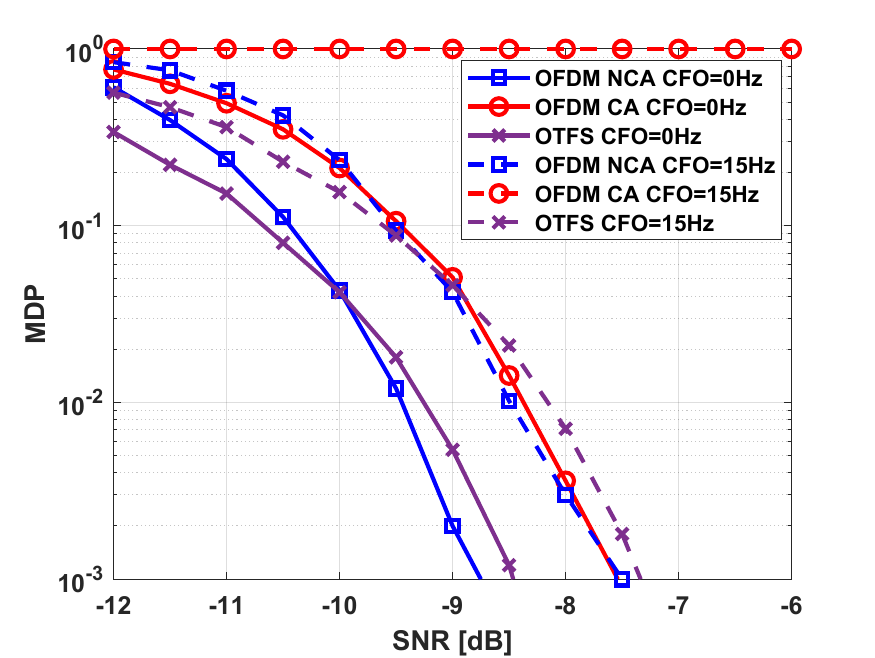}  
\caption{MDP versus SNR for $N_U=1$.}
\label{fig_res}
\end{figure}

\begin{figure}[t]
\centering
\includegraphics[width=0.5\textwidth]{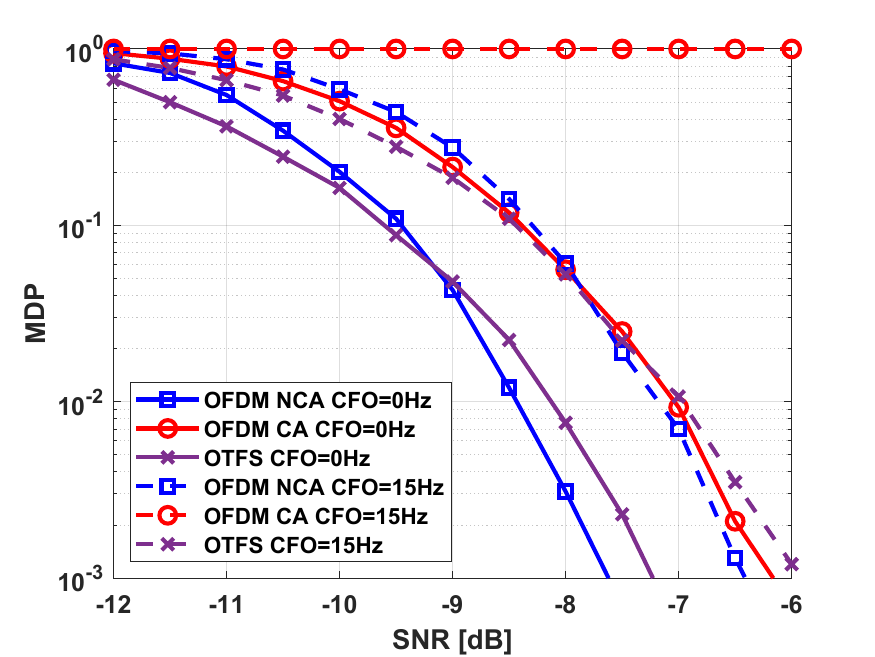}  
\caption{MDP versus SNR for $N_U=2$.}
\label{fig_res2}
\end{figure}

Numerical results are provided in Figure \ref{fig_res} and \ref{fig_res2} for the uncertainties associated with $R_\epsilon=4.3$ km. In such a case, as Figure \ref{fig5} shows, the maximum time and frequency misalignment is respectively given by 49.72 $\mu s$ and $14.46$ kHz. The preamble designs specified in Table \ref{table2} tolerate up to 50 $\mu s$. As far as the delay is concerned, all the contenders are well suited to deal with the impairments that result from an erroneous positioning. In the proposed scenario, the TO is uniformly drawn from $\left[0, \ \ 50 \right]$ $\mu s$, to cover all possible delays. The CFO is fixed to either 0 or 15 kHz, to respectively model ideal and extreme conditions.

The curves depicted in Figure \ref{fig_res} illustrate that OTFS and OFDM NCA achieve comparable performance in terms of preamble detection and delay estimation. The performance gap is limited to a maximum of 0.5 dB. This result indicates that the proposed detector effectively mitigates the adverse effects of fractional delays on OTFS. Moreover, unlike OFDM NCA, OTFS prevents the energy wastage that implies transmitting the CP. As a result, the energy required to transmit the OTFS-based preamble is reduced by a factor $1+(N-1)/N$ compared to its OFDM NCA counterpart. Consequently, the implementation of OFDM NCA comes at the cost of increasing the energy consumption by 2.43 dB. Additionally, it is observed that OFDM CA exhibits poor performance under the evaluated conditions. To understand this result it is essential to note that the size of the detection window determines the number of noise samples that are accumulated. Given that OFDM CA employs the largest detection window, it follows that the noise energy at the detector input will be maximized. 
 
It is obvious that the MDP increases significantly as the CFO approaches $\Delta_f/2$. As shown in Figure \ref{fig_res}, OTFS and OFDM NCA achieve comparable robustness to the CFO. However, it is important to note that OFDM CA is considerably more sensitive to the CFO than the other two schemes. As discussed in Section \ref{secIV}, when OFDM is used as the modulation scheme, the presence of CFO prevents coherent combination of the received sequences for preamble detection. Consequently, OFDM CA fails to support random access under such conditions, as illustrated in Figure \ref{fig_res}.

Figure \ref{fig_res2} shows the MDP in the scenario where two users are simultaneously requesting access in the same time-frequency resources. The analysis focuses on the case where users pick different preambles, thus avoiding collisions. As expected, with the increase in the number of active UEs, the MDP increases as well. This effect arises from the lack of orthogonality among preambles, which are generated from different root sequences. The curves in Figure \ref{fig_res2} indicate that inter-user interference degrades the performance of all schemes to a similar extent.

\section{Conclusions}
\label{secVI}
This work proposes a GNSS-independent random access scheme for NTN, based on modulating identical ZC sequences into the OTFS waveform. The design targets scenarios with large TO and CFO, which are bound to arise in LEO satellite communication systems without GNSS support. By leveraging the structure of OTFS, the proposed scheme enables coherent combination of received signals even in the presence of significant Doppler shifts. Unlike OFDM-based designs, the proposed approach supports joint estimation of fractional and integer delays using a single-step detection algorithm. This eliminates the performance degradation typically caused by error propagation in two-step detection methods. Numerical results confirm that the OTFS-based design achieves performance comparable to state-of-the-art OFDM schemes in terms of MDP and PAPR, while offering clear advantages in spectral confinement and overhead reduction. Notably, the design eliminates the need for a long CP, significantly improving energy efficiency. The scheme also exhibits greater robustness to CFO compared to OFDM with coherent accumulation.


Future work will focus on enhancing the practicality and the robustness of the proposed scheme. First, the current solution assumes that the maximum CFO remains below half the subcarrier spacing. Therefore, new detection algorithms and preamble structures need to be developed to retain the advantages of OTFS, such as a wide delay detection range, while incorporating mechanisms to effectively handle larger CFOs. Second, the structure of the detection algorithm can be enhanced to reduce computational complexity. In parallel, optimizing the preamble sequence to minimize correlation distortion will be crucial for improving detection performance under varying propagation and position uncertainty conditions. Finally, artificial intelligence techniques offer promising opportunities to enhance the design. For instance, machine learning models could assist in estimating delay and Doppler parameters, selecting optimal preamble sequences, or dynamically adapting detection strategies.

\section*{Acknowledgments}
This work has been funded by the 6G-NTN project, which received funding from the Smart Networks and Services Joint Undertaking (SNS JU) under the European Union’s Horizon Europe research and innovation programme under Grant Agreement No 101096479. The views expressed are those of the authors and do not necessarily represent the project. The Commission is not liable for any use that may be made of any of the information contained therein.

Part of the work is funded by the Spanish National project AEl/10.13039/501100011033 SOFIA-SKY (PID2023-1473050B-C32).
\bibliographystyle{IEEEtran}
\bibliography{IEEEabrv,mybibfile}

\begin{IEEEbiographynophoto}{Marius Caus}
received the M.Sc. and Ph.D. (cum laude) degrees in telecommunications engineering from the Universitat Politècnica de
Catalunya (UPC), Barcelona, Spain, in July 2008 and December 2013, respectively. In 2018, he received the two-year postdoctoral Juan de la Cierva Fellowship from the Spanish Government. He is currently a Researcher with CTTC. He has participated in several projects funded by European Commission, European Space Agency, and Spanish Ministry of Science. His main research interests include filter bank-based multicarrier systems, signal processing for communications, and satellite communications.
\end{IEEEbiographynophoto}

\begin{IEEEbiographynophoto}{Musbah Shaat}
received his bachelor’s degree in communication and
control from the IUG university, Palestine, in 2004 and his master’s degree in
communication and electronics engineering from the Jordan University of Science and Technology, Irbid, in 2007. He received his Ph.D. (cum laude) degree in signal theory and communications from the Polytechnic University of Catalonia, Barcelona, Spain, in 2012. He is currently a senior researcher in  Space and Resilient Communications and Systems (SRCOM) research unit at the Centre Tecnològic de Telecomunicacions de Catalunya (CTTC), Barcelona,
Spain. He received the German Academic Exchange Service master’s scholarship in 2005 and was awarded the CTTC Ph.D. fellowship in 2007. He had participated in several national, European, and industrial projects. His main research interests include multicarrier wireless communications, cross-layer optimization, cognitive radio systems, hybrid terrestrial–satellite communications, and cooperative and green communications. 
\end{IEEEbiographynophoto}

\vfill
\end{document}